\documentclass[conference]{IEEEtran}
\IEEEoverridecommandlockouts
\usepackage{cite}
\usepackage{amsmath,amssymb,amsfonts}
\usepackage{algorithmic}
\usepackage{graphicx}
\usepackage{textcomp}
\usepackage{xcolor}
\def\BibTeX{{\rm B\kern-.05em{\sc i\kern-.025em b}\kern-.08em
    T\kern-.1667em\lower.7ex\hbox{E}\kern-.125emX}}
    
    \usepackage{dblfloatfix}

\usepackage{graphicx}
\usepackage{amssymb}
\usepackage{import}
\usepackage{blindtext}
\usepackage{enumitem}
\usepackage{booktabs}
\usepackage{ltablex}
\usepackage{todonotes}
\usepackage{hyperref}
\usepackage{float}
\usepackage{lscape}
\usepackage{longtable}
\usepackage{cleveref}
\usepackage{listings}
\usepackage{multirow}
\usepackage{pgfplots}
\usepackage{changepage}

\usepackage{multirow}
\usepackage{array,graphicx}
\usepackage{arydshln}
\usepackage{caption}

\newcommand*\rotV{\rotatebox{90}}

\pgfplotsset{width=\textwidth,compat=1.8}
\begin{document}
\title{Survey on Tools and Techniques Detecting Microservice API Patterns}
%
%
%

\author{\IEEEauthorblockN{Alexander Bakhtin}
\IEEEauthorblockA{\textit{Tampere University}\\
Tampere, Finland \\
alexander.bakhtin@tuni.fi}
\and

\and
\IEEEauthorblockN{Abdullah Al Maruf}
\IEEEauthorblockA{\textit{Computer Science, Baylor University}\\
Waco, TX, USA \\
maruf\_maruf1@baylor.edu}
\and

\IEEEauthorblockN{Tomas Cerny}
\IEEEauthorblockA{\textit{Computer Science, Baylor University}\\
Waco, TX, USA \\
tomas\_cerny@baylor.edu}

\and
\IEEEauthorblockN{Davide Taibi}
\IEEEauthorblockA{\textit{Tampere University}\\
Tampere, Finland \\
davide.taibi@tuni.fi}
}

\maketitle              
\begin{abstract}
It is well recognized that design patterns improve system development and maintenance in many aspects. While we commonly recognize these patterns in monolithic systems, many patterns emerged for cloud computing, specifically microservices. Unfortunately, while various patterns have been proposed, available quality assessment tools often do not recognize many. This article performs a grey literature review to find and catalog available tools to detect microservice API patterns (MAP). It reasons about mechanisms that can be used to detect these patterns. Furthermore, the results indicate gaps and opportunities for improvements for quality assessment tools. Finally, the reader is provided with a route map to detection techniques that can be used to mine MAPs.

\end{abstract}
\begin{IEEEkeywords}
Microservice API Patterns, Design Patterns, Best Practices, Pattern Matching, Static Analysis, Dynamic Analysis, Pattern Mining Tools
\end{IEEEkeywords}
%
%
%
\section{Introduction}
\label{sec:Introduction}
Design patterns provide a generalized and reusable solution to common software design problems.
They indicate that a system uses best practices. Patterns are well-recognized to improve software quality \cite{iet-sen.2018.5446}. One of the reasons behind such impact is better software comprehension related to software documentation. Detection of patterns is a complex task \cite{Taibi2018closer18,Zimmerman2}. It can constitute static or dynamic system analysis and serves the purpose of quality assurance and quality indices of system design. 



With the era of cloud systems, we must assume new patterns emerge that constitute possibly across multiple self-contained parts of the overall system. In the current context, cloud computing is fueled by the microservice architecture \cite{TheTwelv82:online}. Microservices are small and autonomous services deployed independently, with a single and clearly defined purpose~\cite{Fowler2014,NewmanBook2015}.
We ask what techniques and state-of-the-art tools can detect microservice patterns and the current gaps in this context.

Many challenges need to be taken into account when detecting patterns in cloud-based systems. For instance, the cloud-native development best practices \cite{TheTwelv82:online} suggest separating microservice codebases to enable decentralized evolution. However, current static analysis tools operate on a single codebase \cite{cerny2020access} only. As a result, we cannot detect patterns that span across the whole system by concatenating analysis per codebase. 
This often leads to an alternative direction - dynamic analysis. Dynamic analysis tools can operate on the decentralized perspective \cite{Soldani2021TheT}. However, the dynamic analysis does not reveal a comprehensive detail of the system (e.g., concerning the actual implementation of the service). Quality engineers, architects, and developers might need to know underlying code quality details to improve maintenance and support system comprehension. 

This paper considers a catalog of well-established Microservice-specific API design Patterns (MAP). It reports tools that can detect these patterns and mechanisms for the detection (i.e., static analysis, tracing, log mining). The goal is to identify current gaps in pattern mining and quality assurance automation tools.

To approach this, we adopted a Multivocal Literature Review (MLR) process~\cite{Garousi18a} surveying the systematic gray literature.  
The key motivation for conducting an MLR and therefore including the grey literature is the strong interest of practitioners on the subject, and grey literature content creates a foundation for future research. 

As a result, this work identifies a list of 46 MAPs and 59 pattern mining tools. Out of the 46 patterns, 34 have been addressed by found tools. These 34 MAPs can be discovered by 26 tools out of 59.
Most importantly, we identified gaps in current tools to support MAP pattern identification. We further provide discussion for the reasons behind such a gap and what needs to be addressed to overcome the current hurdles. 


The remainder of this paper is structured as follows. Section~\ref{sec:Background} presents the background and related works.
Section~\ref{sec:design} provides detailed information on the gray literature review process we adopted. Section~\ref{sec:results} reports the results to our RQs. Section~\ref{sec:discussion} discusses implications for practitioners and researchers. Section~\ref{sec:ttv} highlights the threats to validity while finally, Section~\ref{sec:conclusion} draws the conclusions.

\section{Background and Related Work}
\label{sec:Background}

One way to determine software quality is to analyze software for patterns and to verify that it doesn't contain anti-patterns \cite{Taibi2018closer18,Zimmerman2} or to calculate quality metrics such as coupling and cohesion~\cite{Panichella2021}. Different tools have been proposed to perform automatic quality reviews using static analysis of code to operate on pattern detection~\cite{Pigazzini2020}. Pattern mining has been broadly researched \cite{dwivedi2018software,Taibi2018closer18,10.1145/2851613.2851786} and it is a well-established domain, at least for monolithic applications. 

Various microservices patterns have also been identified \cite{TheTwelv82:online,carnell2021spring,Balalaie2018} for various tasks, such as porting from monoliths \cite{Balalaie2018}, supporting resilience \cite{carnell2021spring}, targeting good design practices \cite{Taibi2018closer18, Taibi2020closer2020}, among others. Development frameworks apply these patterns \cite{carnell2021spring} to simplify development.

In this paper, we consider current Microservice API Patterns (MAP further in the text) reported on the API-Pattern website~\cite{Zimmerman,Zimmerman2}. 
The API-Pattern website collects the vast majority of microservices patterns proposed by its creators in peer-reviewed literature~\cite{Zimmerman2,ZimmermanMAP,ZimmermanMAP2,ZimmermanMAP3,ZimmermanMAP4,ZimmermanMAP5,ZimmermanMAP6}. The list of patterns is provided in Table~\ref{tab:patterns}
while the complete description of each pattern can be found on the API-pattern website~\cite{Zimmerman}. We have, however, excluded the pattern {\em ``Annotated Parameter Collection''} as it lacked a detailed description. 

The major difference between monolith and cloud systems in regards to pattern mining, however, is the decentralized codebase, which likely introduces diversity, heterogeneity, and no obvious connection across codebases. Despite initial codebase convention efforts, these get easily lost with evolution and management diversity.

Because of possible diversity across microservices, practitioners often resort to assessing cloud systems through dynamic analysis \cite{app11177856,carnell2021spring}. However, with such direction, we can recognize endpoints and calls but not internal microservice details often needed for pattern detection (since many patterns, including some MAP, have to do with internal implementation as well, e.g. \emph{``Backend Integration''}). 

When we consider other current analysis approaches \cite{app11177856}, we can notice that static analysis of code or mining software repositories involves code parsing and conversion to syntax trees or various graph representations as an intermediate representation. Then it uses these intermediate representations to identify patterns. Other approaches consider log analysis. However, these approaches face difficulty with the non-structured format of log messages. While log clustering can be used, it is very challenging. For this reason, it is common to integrate event tracing, which adds logging statements to calls (i.e., via instrumentation) and collects additional information in log messages, including the originating microservice or the correlation ID to determine distributed transactions and related log messages~\cite{taibi2019}. 

Another approach worth mentioning is program slicing \cite{icisa2020vincent}, which combines log analysis with code analysis. This is accomplished by locating logging statements in the code and identifying logging templates in these statements; these are then matched with log messages found in logs to these code locations \cite{lprof}. For instance, in Lprof \cite{lprof}, the authors used program slicing to profile distributed systems and optimize their performance. They matched log statements with log messages and performed a data-flow analysis of method parameters to identify if these parameters change across call paths. Unchanged parameters identified related log messages and could be used as a correlation ID similar to event tracing. Using this, they could recognize distributed transactions and their frequencies in logs. However, with event tracing (i.e., OpenTelemetry \cite{telemetry}), such tasks become much more simplified and commonly adopted by industry (establishing the correlation ID). Any of these techniques could be used to help with detecting MAPs.












\section{Study Design}
\label{sec:design}
\vspace{-.4em}
This section describes the methods adopted to gather and classify the different tools to detect microservices best practices. 

Since our goal is to map existing tools recommended and adopted by practitioners, we performed a systematic review of the grey literature. 
A review of peer-review literature would be biased toward academic opinions and would not clearly enable us to understand what practitioners can find when looking for such types of tools online. 

In order to investigate the aforementioned goal, we formulated our research questions as: 

\noindent
\textbf{RQ1: Which are the tools available to detect MAPs?
}

This RQ aims to find tools dedicated to mining-specific patterns described in the previous section. Even though some of them could be identified using general-purposes testing tools, the necessity to write tests for the specific patterns makes it unfeasible for the patterns to be adopted in the industry,
while having a set of ready-made tools that could be incorporated into the CD/CI pipeline would facilitate the adoption of best practices and improve code and design quality.

\noindent
\textbf{RQ2: Which MAPs can be detected automatically with tools?
}

In this RQ, we map the tools identified in RQ1 to the patterns they detect.

\noindent
\textbf{RQ3: Which techniques can be used to detect MAPs?
}

In this RQ, we aim to understand whether MAPs can be detected using static or dynamic analysis tools. 
The possibility of detecting the pattern from static analysis tools would enable understanding of the patterns used without running the system. In contrast, the detection based on the dynamic execution of the system and the collection of log traces would allow an understanding of how the system is actually behaving.

\vspace{-1em}
\subsection{The Review Process}
Grey literature Reviews and Multivocal Literature Reviews (MLR) proved to be the best choice for the research method due to the lack of maturity of the subject. MLR includes both academic and grey literature. However, since we are aimed at investigating the word of mouth of practitioners, we will perform a review of the only grey literature. 
The key motivation for the inclusion of grey literature is the strong interest of practitioners in the subject, and grey literature content creates a foundation for future research. 

The process adopted is similar to the MLR, but doesn't include the peer-reviewed literature steps. 

The process we adopted was based on these steps: 
\vspace{-.4em}
\begin{itemize}[label=$\bullet$]
  \item Selection of keywords and search approach
  \item Initial search and creation of an initial pool of sources
  \item Snowballing
  \item Reading through material
  \item Application of inclusion / exclusion criteria
  \item Evaluation of the quality of the grey literature sources
  \item Creation of the final pool of sources
\end{itemize}


\subsection{Literature Search Process}

Since we are interested in finding tools to detect MAPs (Table~\ref{tab:patterns}), we created 53 query search strings.
The search strings were as follows:
\vspace{-.5em}
\begin{itemize}
    \item \texttt{"pattern\_name" api pattern detection tool},
    where \texttt{pattern\_name} was replaced with every
    pattern from table \ref{tab:patterns} (46 total searches, excluding \emph{``Annotated Parameter Collection''} pattern)
    \item \texttt{api security analysis tool}
    \item \texttt{api parameter analysis tool}
    \item \texttt{api parameter discovery tool}
    \item \texttt{api documentation analysis tool}
    \item \texttt{api specification analysis tool}
    \item \texttt{semantic versioning identification tool}
    \item \texttt{microservice api pattern detection tool}
\end{itemize}
\vspace{-.5em}
The latter strings were added because the first parametric search string did not produce many meaningful results, and at the same time, many patterns can be grouped together, as represented by other search strings. We understand that not all groups of patterns are represented by latter search strings, so partial bias is introduced; however, we could not think of search strings targeting other groups of patterns and, as stated before, not using them produced only a handful of results.

\begin{table}[b!]
\footnotesize
\centering
\caption{The list of Microservice API patterns~\cite{Zimmerman}}
\label{tab:patterns}
\begin{tabular}{@{\,}p{14em}@{\,}p{14em}@{\,}}
\textbf{Foundation}         & \textbf{Quality}     \\      
Frontend integration        & API Key                   \\
Backend integration         & Rate Limit                 \\
Public API                  & Rate Plan                  \\
Community API               & Service Level Agreement    \\
Solution Internal API       & Error Report               \\
API Description             & Conditional Request        \\
\textbf{Responsibility}     & Request Bundle             \\
Processing Resource         & Wish List                 \\
Information Holder Resource & Wish Template              \\
Computation Function        & Embedded Entity            \\
State Creation Operation    & \textbf{Structure}             \\     
Retrieval Operation         & Atomic Parameter               \\    
State Transition Operation  & Atomic Parameter List          \\
Operational Data Holder     & Parameter Tree                 \\
Master Data Holder          & Parameter Forest               \\
Reference Data Holder       & Data Element                   \\
Data Transfer Resource      & Id Element                     \\
Link Lookup Resource        & Link Element                   \\                
\textbf{Evolution}          & Metadata Element               \\
Version Identifier          & Annotated Parameter Collection \\
Semantic Versioning         & Context Representation         \\
Two In Production           &\\
Aggressive Obsolescence     &\\
Experimental Preview        &\\
Limited Lifetime Guarantee  &\\
Eternal Lifetime Guarantee  &\\
\end{tabular}

\end{table}

We applied the Search strings to the Google Search\footnote{www.google.com} engine, looking at 10 pages of results per search (excluding ads). The search was done with Incognito browser mode without logging into a personal Google account.
The decision to use 10 pages of results was adopted after an informal piloting of the search, which showed that no relevant results appear on pages 9-10 of the search and that for some patterns (search strings) only a few results (2-3 pages, sometimes not enough even for 1 page) are returned.

Search results consisted of blog posts (including blogs with lists of tools), websites, research papers, Github\footnote{www.github.com} repositories of tools, and Github repositories of lists of tools. It is good to note that StackOverflow\footnote{www.stackoverflow.com} is a popular website for technical peer questions, and it could be expected to appear in the search results, but in reality, it did not. It could have been included in the study separately, but after piloting it, it did not provide meaningful results, so in this paper, we decided to focus only on Google Search.
 
This search was performed between \textbf{10th and 21st of January 2022}.

\subsection{Snowballing}
We applied a backward snowballing to the retrieved literature in the following way:
\begin{itemize}
    \item If the resource extracted from the search is a list of tools (as opposed to a page of just one tool), such as ``Top 10 API security tools in 2021''\footnote{This is a made-up example}, then we checked all of the referenced tools (and potentially other referenced lists)
    \item If the resource is a research paper, we checked if the paper cites other algorithms that fit our criteria and included them as well
\end{itemize}

\subsection{Application of inclusion / exclusion criteria}

Based on MLR guidelines~\cite{Garousi18a}, we defined our inclusion criteria:
\begin{itemize}
    \item For pages of tools: the tool description directly contains one of the MAPs from \ref{tab:patterns}
    
    \item For research papers: the paper proposes a new algorithm/tool to detect a pattern whose description in the abstract is similar to studied  patterns
\end{itemize}

Moreover, we defined our exclusion criteria as: 

\begin{itemize}[label=$\bullet$]
  \item Exclusion criterion 1: Non-English results 
  \item Exclusion criterion 2: Duplicated result
  \item Exclusion criterion 3: (for papers) The paper proposes a new algorithm/tool to detect a pattern; however, no source code for the tool is provided
  \item Exclusion criterion 4: (for commercial tools) The tool has no public documentation of functionality available
  \item Exclusion criterion 5: (for tools) The tool is an all-purpose security tool claiming to, e.g., ``identify over 1000 different vulnerabilities'', i.e., no particular reference to one of the MAP is given.
  \item Exclusion criterion 6: (for tools) The tool cannot automatically perform the analysis but requires the programmers to configure it and adapt a certain workflow first (e.g., configure the tool to detect certain words in commit messages and then use them in work), so that it cannot be used to retrospectively analyze the history of an existing project.
\end{itemize}


\subsection{Evaluation of the quality and credibility of sources}

In order to evaluate the credibility and quality of the selected grey literature sources and to decide whether to include a grey literature source or not, we extended and applied the quality criteria proposed by Garousi et al. ~\cite{Garousi18a}, considering the authority of the producer, the methods applied, objectivity, date, novelty, impact, and outlet control. We adopted the same evaluation sheet adopted by Peltonen et al. ~\cite{Peltonen2020}.

Two authors assessed each source using the aforementioned criteria, with a binary or 3-point Likert scale, depending on the criteria themselves. In case of disagreement, we discussed the evaluation with the third author, which helped to provide the final assessment.

We finally calculated the average of the scores and rejected grey literature sources that scored lower than 0.5 on a scale that ranges from 0 to 1.

\subsection{Creation of final pool of sources}
\label{subsec:FinalPoolOfSources}
Originally, 45 different resources were identified as relevant.
After performing snowballing on papers and lists
and filtering all papers and tools through the exclusion criteria,
we had a list of 59 tools.
\textbf{}

\subsection{Data Extraction}

In order to obtain the list of tools to detect MAPs (\textbf{RQ1}) we extracted the tool names from the selected sources.


To understand which pattern is detected by the tools (\textbf{RQ2}), firstly, we read through the tool documentation. There are two sorts of documents available: a dedicated website and a README file. Few tools have both. 
After reading the documentation, we investigated whether this tool recognizes any MAP and whether it is possible to obtain information about those patterns using this tool. We couldn't locate any tool that recognizes MAP on its own because these MAPs are not commonly well-established yet. However, a few tools disclose data that can be utilized to develop an insight into MAPs. 

We examined each tool by reviewing its documentation to see if it exposed any information that may be utilized to detect any MAP. When we locate a tool that can detect a pattern, we map that tool to that pattern. 

Lastly, to understand which technique can be adapted to detect MAPs (\textbf{RQ3}),
We manually analyzed each pattern, including those not discovered by tools, and mapped each pattern to the possible technique. For the techniques, we considered static and dynamic analysis. For static analysis, we considered plain code analysis, operation with call graphs, which might require more advanced algorithms, and mining software repositories which may include additional information. For dynamic analysis, we considered application log analysis considering rich logging in the system code (an ideal case) and event logs (i.e., received from instrumentation with correlation ID \cite{carnell2021spring}). We did not consider program slicing.

The first two authors both independently examined and mapped each pattern to the various techniques using their own reasoning. After that, the differing viewpoints were resolved by consulting with each other as well as the other authors.




\section{Results}
\label{sec:results}
This section reports the results we obtained following the research methodology highlighted in Section 4.

As for the tools available to detect MAPs (\textbf{RQ1}), we identified 59 tools from 45 sources. 
Tables \ref{tab:rq1oss} and \ref{tab:rq1comm} list the tools retrieved (Open Source and Commercial, respectively), together with their URL,
license (for Open Source tools), languages supported for analysis, and date of the last update (for Open Source tools). As for the Licenses adopted by OSS projects, the license indicated in their repository is stated, such as MIT, Apache, etc.; `OSS' refers to tools whose source code is openly available, but no license is added to the repository; `N/A (Free)' refers to tools that are available for free (e.g., as a web service), but not in Open Source form, and thus might be subject to a custom license as well.
Different tools can analyze MAPs from the perspective of different programming languages. Some tools support several languages. Other tools scan git commits to perform the analysis, and thus applicable to any language. While other tools either access the APIs under analysis using the provided endpoints or analyze API specifications in OpenAPI format, thus the language of implementation doesn't matter. The language reported in Table~\ref{tab:rq1oss} as  `Any' refers to tools that parse source code in a language-agnostic manner and thus apply to any language. 

\begin{table*}[]
\caption{OSS tools to detect Microservice API Patterns (RQ1)}
\label{tab:rq1oss}
\centering
\begin{tabular}{l|lllll}
\textbf{} & \textbf{}              & \textbf{}   & \textbf{} & \textbf{Language} & \textbf{} \\
\textbf{ID} & \textbf{Tool}              & \textbf{URL}   & \textbf{License} & \textbf{supported} & \textbf{Updated} 

\\ \hline
T1               & API security tools audit   & https://apisecurity.io/tools/audit/ & N/A (Free)       & REST API                 & N/A (Active)     \\
T2               & apidiff                    & bit.ly/3J0FyZN & MIT              & Java                     & 31.10.2021       \\
T3               & Arjun                      & bit.ly/3uFZoWv & GPL-3.0          & REST API                 & 29.08.2021       \\
T4               & Astra                      & bit.ly/3ontBp4 & Apache-2.0       & REST API                 & 05.04.2019       \\
T5               & Brakeman                   & bit.ly/3LbQtSo & MIT              & Ruby                     & 30.01.2022       \\
T6               & Coala                      & bit.ly/3AXgKiw & AGPL-3.0         & Any                      & 11.06.2021       \\
T7               & code2flow                  & bit.ly/34w2MYM & MIT              & Many (4)                 & 27.12.2021       \\
T8               & git-secret                 & git-secret.io  & MIT              & git                      & 01.02.2022       \\
T9               & git-semv                   & bit.ly/34yi1jM & MIT              & git                      & 17.06.2021       \\
T10              & GitVersion                 & bit.ly/3AVeUi6 & MIT              & git                      & 31.01.2022       \\
T11              & go-semrel-gitlab           & bit.ly/3upDINT & MIT              & git                      & 27.10.2019       \\
T12              & GraphQL FBC-CLI            & bit.ly/333l57q & OSS              & GraphQL                  & 06.06.2018       \\
T13              & Hikaku                     & bit.ly/3Hubkhx & Apache-2.0       & REST-API                 & 19.08.2021       \\
T14              & jgitver                    & bit.ly/3glahV5 & Apache-2.0       & Java                     & 30.03.2021       \\
T15              & Kiterunner                 & bit.ly/3J77eML & AGPL-3.0         & REST API                 & 10.05.2021       \\
T16              & LAPD                       & bit.ly/3gl6FlZ & N/A (Free)       & Java                     & 07.09.2013       \\
T17              & magento-semver             & bit.ly/3gnWygi & OSL-3.0          & N/A                      & 19.01.2022       \\
T18              & microservices-antipatterns & bit.ly/3GwBwGJ & Apache-2.0       & Python                   & 17.12.2019       \\
T19              & modver                     & bit.ly/32WbY8k & MIT              & Go                       & 16.01.2022       \\
T20              & Mondrian                   & bit.ly/3rqHkNZ & OSS              & PHP                      & 16.09.2014       \\
T21              & MSA-nose                   & bit.ly/3sgODXF & OSS              & Java                     & 12.04.2021       \\
T22              & next-ver                   & bit.ly/3B5EWzq & MIT              & git                      & 09.02.2018       \\
T23              & NodeJS scan                & bit.ly/3uqxHkm & GPL-3.0          & Node.js                  & 31.01.2022       \\
T24              & NoRegrets                  & bit.ly/3JjDAnF & Apache-2.0       & JavaScript               & 02.07.2019       \\
T25              & OpenAPI diff               & bit.ly/3oncymY & MIT              & REST API                 & 29.08.2017       \\
T26              & OpenAPI spec validator     & bit.ly/3sogwxc & Apache-2.0       & REST API                 & 28.01.2022       \\
T27              & openapi-lint               & bit.ly/3oqUJDA & BSD-3            & REST API                 & 12.08.2020       \\
T28              & openapilint                & bit.ly/3J45Lqs & MIT              & REST API                 & 13.05.2019       \\
T29              & oval                       & bit.ly/3J466te & MIT              & REST API                 & 26.09.2018       \\
T30              & pact                       & bit.ly/3gp77Q8 & MIT              & REST-API                 & 03.02.2022       \\
T31              & paramspider                & bit.ly/3sf6JZZ & GPL-3.0          & REST API                 & 12.09.2021       \\
T32              & Prometheus                 & prometheus.io  & Apache-2.0       & Many (5)                 & 02.02.2022       \\
T33              & prospector                 & bit.ly/3oo4Zg7 & GPL-2.0          & Python                   & 01.02.2022       \\
T34              & Public API changes         & bit.ly/3gjOKMF & Unlicense        & C\#                      & 05.11.2017       \\
T35              & pycallgraph                & bit.ly/3uqyZfc & GPL-2.0          & Python                   & 28.02.2018       \\
T36              & Pyramid OpenAPI3           & bit.ly/3uqbZg5 & MIT              & Python                   & 07.12.2021       \\
T37              & pyramid-swagger            & bit.ly/3sfLKqb & BSD-3            & Python                   & 30.03.2020       \\
T38              & Python Semantic Release    & bit.ly/3oo3uyg & MIT              & Python                   & 31.01.2022       \\
T39              & REST API Antip. Inspect.   & bit.ly/3GkLhrG & MIT              & REST API                 & 31.03.2021       \\
T40              & schaapi                    & bit.ly/3J3BcB7 & MIT              & Java                     & 11.02.2019       \\
T41              & secret-detection           & bit.ly/3rqsM0N & OSS              & Any                      & 03.08.2020       \\
T42              & semantic-release           & bit.ly/3usLkPQ & MIT              & git                      & 18.01.2022       \\
T43              & semantic-versioning-anal.  & bit.ly/3HsvgBn & MIT              & .NET                     & 03.11.2021       \\
T44              & semver-config              & bit.ly/3GnLpGN & OSS              & git                      & 26.09.2019       \\
T45              & semverbot                  & bit.ly/3GteoZI & MPL-2.0          & git                      & 03.01.2022       \\
T46              & Speccy                     & bit.ly/3uknqGe & MIT              & REST API                 & 02.10.2019       \\
T47              & Spectral                   & bit.ly/3gnIeo3 & Apache-2.0       & REST API                 & 03.02.2022       \\
T48              & SpotBugs                   & bit.ly/3glbvQj & LGPL-2.1         & Java                     & 29.01.2022       \\
T49              & Standard Version           & bit.ly/3uqgx64 & ISC              & Node.js                  & 01.01.2022       \\
T50              & Vulture                    & bit.ly/3Gq9hd2 & MIT              & Python                   & 03.01.2022       \\
T51              & wFuzz                      & bit.ly/3opt5Hg & GPL-2.0          & REST API                 & 28.11.2020       \\
T52              & Zally                      & bit.ly/34lLHkq & MIT              & REST API                 & 14.01.2022
\end{tabular}%

\end{table*}
\begin{table*}[]
\caption{Commercial tools to detect Microservice API Patterns (RQ1)}
\label{tab:rq1comm}
\centering
\begin{tabular}{l|lll}
\textbf{ID}&\textbf{Tool}           & \textbf{URL}   & \textbf{Supported Languages} \\ 
\hline
T53&Acunetix                & acunetix.com   & Web                      \\
T54&CheckMarx               & checkmarx.com  & Many (20)                \\
T55&CodeClimate             & bit.ly/3GrqB12 & Many (11)                \\
T56&Data Theorem API Secure & bit.ly/34ASaYM & N/A                      \\
T57&Dynatrace               & dynatrace.com  & Many                     \\
T58&SonarQube*               & bit.ly/3GryR15 & Many (29)                \\
T59&Synopsys                & bit.ly/3uqjhRd & REST API                
\\\hline
\multicolumn{4}{l}{*Dual-licensed, available both as commercial and open source (GPL).}
\\\hline
\end{tabular}
\end{table*}

 There are 7 Commercial tools and 52 OSS tools to provide some summary statistics.
 
 When it comes to OSS licenses, 33 of tools (56\% of total tools) are using permissive licenses (Apache, MIT, BSD, etc.), and 5 more use no license at all, while 7 tools (11\%) use `copyleft' licenses (different versions of GPL license).
 
 In terms of languages, 16 tools (27\%) are written in Python, 11 tools (19\%) are written in JavaScript (some of them in TypeScript), Java and Go have 6 tools each (10\% each); other represented languages are C++/C\#, Kotlin, PHP and shell scripting (bash).
 
 Supported languages/platforms involve 8 tools analyzing commits in Git (14\%), another 8 tools targeting Python as the only language (13\%), 6 tools (10\%) targeting Java, and another 6 tools (10\%) supporting several languages. Also, 19 tools (32\%) target REST APIs directly, with 9 analyzing specifications and 10 using the endpoints dynamically.
 
 Out of 52 OSS tools, 32 have been updated at least once since January 2021. Furthermore, 19 have been updated already in 2022.
 
 It should also be noted that some tools are research prototypes (T2, T16, T18, T21), some grew out of research prototypes (T48), while many are projects done by hobbyists (T19, T34, T37 to name a few), so their quality and applicability to up-to-data languages and frameworks could be limited. The scope of this paper is simply to identify existence of \emph{some} tools to address MAPs and see the pattern coverage, assessing the actual quality and usefulness of the tools is a different, much more complicated endeavour. 
 
As for the MAPs that can be detected automatically with tools (\textbf{RQ2}), we found 26 tools that expose information about the tools. These tools are listed as RQ2 column in Table \ref{tab:rq3}. Our table shows that found tools detect a subset of all patterns, and a combination of tools is necessary to address broader coverage. For example, the pattern `API Description' is detected by 9 tools in the Foundation category, while no tool targets the other 5 patterns in this category. Two tools, code2flow, and pycallgraph, which are both based on call graphs, can identify all of the patterns in the Responsibility category. Hikaku is a tool that can be used to detect all of the patterns in the Structure category. All of the patterns in this category, with the exception of `Context representation,' can be discovered with four or more tools. There are three patterns in the quality section that no tool can detect. Rate Plan, Service Level Agreement, and Wish Template are examples of these patterns. At least one tool detects the rest of the patterns. Three patterns are not detected by any tools in the Evolution category, whereas four patterns are detected by at least three tools.



The most promising techniques to detect MAPs (\textbf{RQ3}) stem from the static analysis involving source code analysis as detailed in Table \ref{tab:rq3}. Most of the tools we found use static analysis. In particular, the static analysis might need to determine call graphs to detect certain patterns, and also, the detection process can use mining software repositories. Nevertheless, given the API level, dynamic analysis can also determine a large number of patterns. While application log analysis is one option, it is a challenging option dependent on the level of logging. It is more convenient to use event logs resulting from recent cloud-native frameworks and infrastructure advancements. All events are centralized and aggregated by the occurrence time, with correlation ID indicating message dependencies. 



\begin{table*}
	\scriptsize
	\centering
	\caption{Possible Techniques for detecting microservice patterns}
	\label{tab:rq3}
	\begin{tabular}{|l|c|ccc|cc|}
		\hline
		& \textbf{RQ2}                 & \multicolumn{5}{c}{\textbf{RQ3}}    \\ \cline{3-7}
		Patterns   &  Detected by tool (ID)                  & \multicolumn{3}{c|}{Static} & \multicolumn{2}{c}{Dynamic}     \\
		&                     & \multicolumn{3}{c|}{Analysis} & \multicolumn{2}{c}{Analysis}     \\
		\hline

		&&
		\rotV{Call Graph} &
		\rotV{Source Code} &
		\rotV{Repository} &
		\rotV{Event Log} &
		\rotV{Application log}
		\\

		\hline
		\multicolumn{7}{|c|}{\textbf{Foundation}}                                                                                                 \\
		\hline
		Frontend integration        &                                                & \checkmark & \checkmark &            & \checkmark & \checkmark \\
		\hline
		Backend integration         &                                                & \checkmark & \checkmark &            &            &            \\
		\hline
		Public API                  &                                                &            &            &            &            &            \\
		\hline
		Community API               &                                                &            &            &            &            &            \\
		\hline
		Solution Internal API       &                                                &            &            &            &            &            \\
		\hline
		API Description             & T25, T26, T27, T28, T29, T36, T37, T46, T52    &            & \checkmark &            &            &            \\
		\hline
		\multicolumn{7}{|c|}{\textbf{Responsibility}}                                                                                             \\
		\hline
		Processing resource         & T7, T35                                        & \checkmark & \checkmark &            & \checkmark & \checkmark \\
		\hline
		Information holder resource & T7, T35                                        & \checkmark & \checkmark &            & \checkmark & \checkmark \\
		\hline
		Computation function        & T7, T35                                        & \checkmark & \checkmark &            & \checkmark & \checkmark \\
		\hline
		State recreation operation  & T7, T35                                        & \checkmark & \checkmark &            & \checkmark & \checkmark \\
		\hline
		Retrieval operation         & T7, T35                                        & \checkmark & \checkmark &            & \checkmark & \checkmark \\
		\hline
		State transition operation  & T7, T35                                        & \checkmark & \checkmark &            & \checkmark & \checkmark \\
		\hline
		Operational Data Holder     & T7, T35                                        & \checkmark & \checkmark &            & \checkmark & \checkmark \\
		\hline
		Master data holder          & T7, T35                                        & \checkmark & \checkmark &            & \checkmark & \checkmark \\
		\hline
		Reference data holder       & T7, T35                                        & \checkmark & \checkmark &            & \checkmark & \checkmark \\
		\hline
		Data transfer resource      & T7, T35                                        & \checkmark & \checkmark &            & \checkmark & \checkmark \\
		\hline
		Link Lookup resource        & T7, T35                                        & \checkmark & \checkmark &            & \checkmark & \checkmark \\
		\hline
		\multicolumn{7}{|c|}{\textbf{Structure}}                                                                                                  \\
		\hline
		Atomic parameter            & T3, T6, T13, T31, T36, T37                     &            & \checkmark &            &            &            \\
		\hline
		Atomic parameter list       & T3, T6, T13, T31, T36, T37                     &            & \checkmark &            &            &            \\
		\hline
		Parameter tree              & T3, T6, T13, T31, T36, T37                     &            & \checkmark &            &            &            \\
		\hline
		Parameter forest            & T3, T6, T13, T31, T36, T37                     &            & \checkmark &            &            &            \\
		\hline
		Data element                & T6, T13, T31, T36, T37                         &            & \checkmark &            &            &            \\
		\hline
		Id element                  & T3, T6, T13, T31, T36, T37                     &            & \checkmark &            &            &            \\
		\hline
		Link element                & T6, T13, T31, T36, T37                         &            & \checkmark &            &            &            \\
		\hline
		Metadata element            & T13, T31, T36, T37                             &            & \checkmark &            &            &            \\
		\hline
		Context representation      & T13                                            &            & \checkmark &            &            &            \\
		\hline
		Pagination                  & T13, T31, T36, T37                             &            & \checkmark &            & \checkmark & \checkmark \\
		\hline
		\multicolumn{7}{|c|}{\textbf{Quality}}                                                                                                    \\
		\hline
		API key                     & T13, T36, T37                                  &            & \checkmark &            & \checkmark & \checkmark \\
		\hline
		Rate limit                  & T36, T37, T53                                  &            & \checkmark &            & \checkmark & \checkmark \\
		\hline
		Rate plan                   &                                                &            &            &            &            &            \\
		\hline
		Service level agreement     &                                                &            &            &            &            &            \\
		\hline
		Error report                & T13, T36, T37                                  & \checkmark & \checkmark &            & \checkmark & \checkmark \\
		\hline
		Conditional request         & T13, T36, T37                                  & \checkmark & \checkmark &            & \checkmark & \checkmark \\
		\hline
		Request bundle              & T13                                            & \checkmark & \checkmark &            & \checkmark & \checkmark \\
		\hline
		Wish list                   & T36, T37                                       & \checkmark & \checkmark &            & \checkmark & \checkmark \\
		\hline
		Wish template               &                                                & \checkmark & \checkmark &            & \checkmark & \checkmark \\
		\hline
		Embedded entity             & T13                                            & \checkmark & \checkmark &            & \checkmark & \checkmark \\
		\hline
		Linked information holder   & T13                                            & \checkmark & \checkmark &            & \checkmark & \checkmark \\
		\hline
		\multicolumn{7}{|c|}{\textbf{Evolution}}                                                                                                  \\
		\hline
		Version identifier          & T6, T10, T13, T45                              &            & \checkmark & \checkmark & \checkmark &            \\
		\hline
		Semantic versioning         & T2, T9, T13, T19, T38, T40, T42, T43, T44, T45 &            & \checkmark & \checkmark & \checkmark &            \\
		\hline
		Two in production           &                                                &            & \checkmark & \checkmark & \checkmark &            \\
		\hline
		Aggresive obsolesense       &                                                & \checkmark & \checkmark & \checkmark & \checkmark &            \\
		\hline
		Experimental preview        &                                                &            & \checkmark & \checkmark & \checkmark &            \\
		\hline
		Limited lifetime guarantee  & T2, T13, T40                                   &            & \checkmark &            &            &            \\
		\hline
		Eternal lifetime guarantee  & T2, T13, T40                                   &            & \checkmark &            &            &            \\
		\hline
	\end{tabular}

\end{table*}

\section{Discussion}
\label{sec:discussion}
It is interesting to note that the vast majority of MAPs can be detected by tools, even if there are no tools that analyze them all. 
Foundation patterns are the only group of MAPs where no tools implemented their detection, except for the "API description." However, few other patterns could be technically detected, and therefore tools could implement them. Other pattern categories have rather reasonable coverage by tools.


Regarding {\bf RQ1}, we found 59 tools available for the detection task. We listed these tools in Tables \ref{tab:rq1oss} and \ref{tab:rq1comm}. These tables also divided the tools based on open-source availability. We further categorized these tools based on patterns they identify in Table \ref{tab:rq3}. However, not all tools had available information on which patterns they could detect.

With regards to {\bf RQ2}, almost no tool identifies the patterns directly. The extracted results must be post-interpreted by users to identify these patterns from the provided information. For example, tools discovering the pattern 'Semantic Versioning Identifier' do not simply tell if the project has followed the SemVer specification\footnote{www.semver.org}, but instead tell the correct identifier (increment) based on changes in the source, and it is up to the developer/researcher to compare it to the one actually used. This is a missing step to better terminology unification, establishment, and automation in the domain, which some practitioners could desire.
Still, there are notable gaps and improvement opportunities. Despite 59 tools found, there is no outstanding tool with respect to the detection coverage of a number of patterns. Tools must be combined to address different types of patterns. Table \ref{tab:rq3} outlines identified gaps that quality assurance tools should fill to provide better quality measures through integration into a single solution.
Some conventional API testing tools could be used to detect specific patterns, but we excluded such tools as they need explicit scripting. As an example, in Postman \cite{postman}, it is possible to extract request/response headers from HTTP calls, and from the headers, users can write tests to see if it contains API Key or Version information, etc. 

 
Related to {\bf RQ3}, identified tools are predominantly based on static code analysis, and more advanced techniques might be necessary to detect some patterns as depicted by Table \ref{tab:rq3}. This table also shows that some patterns that could be detected by the techniques we identified are not yet recognized by the pool of tools we found, which opens opportunities. Some of the patterns currently require manual input to be determined. However, this opens questions about whether other techniques could be considered to address these patterns. 
For instance, \emph{``Rate Plan''} and \emph{``SLA''} are about the legal use of the API, and perhaps organizational policies need to be taken into account. Still, these policies are not in a machine-readable format, and there is no guarantee these are enforced; thus, more advanced mechanisms would need to be considered to allow combinations of static or dynamic analysis with organizational policies.



The results of this work can be useful for researchers that can investigate different techniques for detecting patterns. This paper's outcomes could also be useful to practitioners who can access the list of tools that automatically detect patterns and eventually integrate them in continuous quality control models~\cite{Lenarduzzi2017ECISME}. Finally, results might be beneficial to tool providers that might extend their solutions to detect a large number of patterns or integrate them into DevOps pipelines~\cite{Taibi2019CCSS}.


\section{Threats  to Validity}
\label{sec:ttv}
Systematic reviews and surveys often suffer from several threats to validity. We discuss the threats 
considering construct, internal, external, and conclusions validity.

The construct validity is meant to consider the research questions within the investigated area.
Our queries are motivated by gaps in related works. The search terms combined established terms and pattern names commonly recognized in the community and domain of this work.
We addressed a possible threat of omitting relevant research from our review by experimenting with several other search strings identifying related work. The analyzed sample considered grey literature articles to ensure the up-to-date perspective of practitioners. Furthermore, our evidence search was often limited to an abstract overview of sources, which could miss relevant work.

Internal validity involves methods to study and analyze data (e.g., the types of bias involved).
One potential threat is related to inclusion and exclusion, a process that includes metadata. Besides, our bias could affect the filtering. Multiple authors performed this search, with primary authors assigned to particular patterns and secondary authors spot-checking. Apart from the filtering process, we performed tool localization by name, documentation, and repository identification if available. Our approach to taxonomy is a result of our discussions of interpreted results and represents our view on the identified literature and domain.

External validity is related to knowledge generalization. This survey interprets and categorizes works we gathered from established scientific channels and grey literature along with our experience related to the field. We could have missed related work on specific patterns or related tools because of our selection of search strings, part of which target specific patterns or groups of patterns. However, even with a subset of identified tools and approaches, we would likely have identified common techniques applied for pattern mining and provided an overview of what has been accomplished and which gaps remain to be addressed.

The conclusions resulting from several brainstorming sessions were independently settled and agreed on by all authors.
To address the validity of the conclusions, we involved multiple authors in this study with diverse backgrounds, all discussing the outcomes in the context of extracted and synthesized information.

\section{Conclusion}
\label{sec:conclusion}
This paper considered microservice API patterns (MAP) and their recognition by available quality assurance tools. It provides a practical road map to what tools and open-source exits to detect these patterns and where the gaps remain. We have identified 59 tools that address 34 MAPs out of 46 identified. We did not find a specific tool that would surpass others, and thus a combination of tools is necessary to cover a broader spectrum of MAPs. Yet, not complete coverage exists considering our search results. This gap represents an opportunity for the community to combine and integrate efforts to provide better quality assurance tools needed in this mainstream field. Furthermore, considering that we identified many open source projects, their integration is a logical next step to developing a more advanced microservice infrastructure for quality assurance.

\section*{Acknowledgments}

This material was supported by the ADOMS Grant awarded by the Ulla Tuominen Foundation (Finland), National Science Foundation under Grant No. 1854049 and a grant from Red Hat Research \url{https://research.redhat.com}.

%
%
 \bibliographystyle{IEEEtran}
  \bibliography{bibliography}

\end{document}